\documentclass{moriond}

\def\Journal#1#2#3#4{{#1} {\bf #2}, #3 (#4)}

\def\be{\begin{equation}}
\def\ee{\end{equation}}
\def\bea{\begin{eqnarray}}
\def\eea{\end{eqnarray}}

\newcommand{\Photo}{}

\begin{document}
\vspace*{4cm}
\title{GENERALIZED DARK MATTER MODEL WITH THE EUCLID SATELLITE}

\author{ I. TUTUSAUS, B. LAMINE, AND A. BLANCHARD }

\address{IRAP, Universit\'{e} de Toulouse, CNRS, CNES, UPS, (Toulouse), France}

\maketitle\abstracts{The concordance model in cosmology, $\Lambda$CDM,
  is able to fit the main cosmological observations with a high level
  of accuracy. However, around 95\% of the energy content of the
  Universe within this framework remains still unknown. In this work
  we focus on the dark matter component and we investigate the generalized dark matter (GDM) model, which
  allows for non-pressure-less dark matter and a non-vanishing sound
  speed and viscosity. We first focus on current observations, showing
  that GDM could alleviate the tension between cosmic microwave
  background and weak lensing observations. We then investigate the ability
  of the photometric Euclid survey (photometric galaxy clustering,
  weak lensing, and their cross-correlations) to constrain the nature of dark
  matter. We conclude that Euclid will provide us with very good
  constraints on GDM, enabling us to better understand the nature of
  this fluid, but a non-linear recipe adapted to GDM  is clearly needed
  in order to correct for non-linearities and get reliable results
  down to small scales.}

\section{Introduction}
$\Lambda$CDM has become the concordance model in cosmology thanks
to its ability to fit the main cosmological observations\,\cite{Planck2016}. It  is mainly
characterized by its dark sector, composed of pressure-less, non-interacting cold dark
matter (CDM) and a cosmological constant $\Lambda$. %In particular, these components
%are usually considered perfect fluids and modeled with no pressure
%(for cold dark matter) and an equation of state parameter $w=-1$ for
%the cosmological constant. 
However, their nature remain still
unknown. In this work we
focus only on the dark matter component of the Universe and we follow
a phenomenological approach to go beyond the standard model. In particular, we use the generalized
dark matter (GDM) model, first proposed by Hu\,\cite{Hu1998}, to constrain dark
matter properties in the linear regime. We first present the
theoretical framework of the GDM model in Sec.\,\ref{sec_2}. We then
show the constraints on this model using current observations in
Sec.\,\ref{sec_3}, and we finally show the expected precision of the
photometric Euclid
survey on the GDM model parameters in Sec.\,\ref{sec_4}, before
finishing with the conclusions in Sec.\,\ref{conclusions}.

\section{Theoretical framework}\label{sec_2}

In this work we assume that dark matter is only coupled to the visible
sector through gravitational interaction, so we assume that the dark
matter energy-momentum tensor is conserved. This implies that all kind of dark matter
components can be covered by the standard conservation equations for a
general matter source\,\cite{Ma1995}. The
dark matter energy density $\rho$ evolves as
$\dot{\rho}+3H(1+w)\rho=0$, where the over-dot stands for the derivative with respect to conformal
time, and $H\equiv \dot{a}/a$ is the conformal Hubble
parameter. In this work we focus on the scalar modes, neglecting
vector and tensor perturbations. Therefore, a conserved energy-momentum tensor must satisfy\,\cite{Ma1995} (at a
linear level of perturbations and in the synchronous gauge)

{\small \begin{equation}
\dot{\delta}+(1+w)\left(\theta+\frac{\dot{h}}{2}\right)+3H\left(\frac{\delta
    p}{\delta\rho}-w\right)=0\,\hspace{5pt}{\rm
  and}\hspace{5pt}\dot{\theta}+H(1-3w)\theta+\frac{\dot{w}}{1+w}\theta-\frac{\delta p/\delta\rho}{1+w}k^2\delta+k^2\sigma =0\,,\label{eq_GDM_pert2}
\end{equation}}
where $w$, $\delta$, and $\theta$ stand for the fluid equation of
state parameter, its density fluctuation
and the divergence of its velocity, respectively. $\delta p$
represents the pressure perturbation, and $\sigma$ corresponds to the
anisotropic stress. Provided Eq.\,(\ref{eq_GDM_pert2}), the GDM model
is specified by the dark matter
equation of state parameter $w$, and relations between $\delta p$ and $\sigma$ to the
dynamically evolving variables $\delta,\,\theta$. We consider
non-relativistic dark matter (it can allow for the formation of
galaxies) with the so-called
$c_{\rm vis}$ parametrization\,\cite{Hu1998}, where $\delta p$ is related to $\delta$ and $\theta$ through the
rest-frame sound speed $c_s$, and $\sigma$
evolves according to:

{\small \begin{equation}
\delta p=c_s^2\delta \rho
-\dot{\rho}(c_s^2-c_a^2)\theta/k^2\,\hspace{5pt}{\rm and}\hspace{5pt}\dot{\sigma}+3H\frac{c_a^2}{w}\sigma=\frac{4}{3}\frac{c_{\rm vis}^2}{1+w}(2\theta+\dot{h}+6\dot{\eta})\,,
\end{equation}}
where the adiabatic sound speed is $c_a^2\equiv
(w\rho)^{\dot{}}/\dot{\rho}$, $c_{\rm vis}^2$ is a new viscosity parameter, and $h$ and
$\eta$ are the synchronous metric perturbations. We fix $c_{\rm vis}^2=0$, for
simplicity, and we consider only the equation of state parameter $w$
and the sound velocity $c_s^2$ as constant parameters for the GDM model.

\section{Current constraints}\label{sec_3}

%\subsection{Method and data}
We first need a Boltzmann code to compute the power
spectrum for the GDM model. In this work we use the \texttt{CLASS} code\,\cite{class2}. It already includes a parametrization for the dark energy
fluid with a constant equation of state parameter and a constant sound
velocity\,\cite{class4}, so we use this parametrization as GDM,
while we keep a cosmological constant for the dark energy
contribution, and a negligible fraction of cold
dark matter. Notice that the perturbations computed for this
fluid must be added to the total matter perturbations, which is not
the case in the default version of \texttt{CLASS}, since the fluid is supposed
to behave as dark energy. We then investigate the constraints on the cosmological parameters
using a Markov chain Monte Carlo approach, with the
Metropolis-Hastings algorithm, implemented in the parameter inference
Monte Python code\,\cite{Audren2013}. We use the Gelman-Rubin
test\,\cite{Gelman1992}, requiring $1-R<0.015$ for all parameters, to
consider that the chains have converged. We consider the 6 baseline
parameters for $\Lambda$CDM that can be seen in Table\,\ref{table1}, plus $w$ and $c_s^2$ for GDM. We further consider
two massless neutrinos and a massive one with mass 0.06\,eV, keeping
the value of the effective number of neutrino-like relativistic
degrees of freedom $N_{\rm eff}=3.046$. The constraints on the parameters are obtained using CMB data (the 2015 Planck CMB likelihoods\,\cite{PlanckGDM1,PlanckGDM2}), baryon acoustic
oscillations (BAO)
measurements  from BOSS\,\cite{Anderson2014},
6dFGS\,\cite{Beutler2011}, and SDSS\,\cite{Ross2015}, and type Ia
supernovae (SNIa) data from the
joint light-curve analysis (JLA)\,\cite{Betoule2014}. For some runs we
include weak lensing (WL) data from the CFHTLenS survey\,\cite{Heymans2013}
to check whether the tension between WL and CMB data is alleviated
when considering GDM.

\begin{figure}
\begin{center}
\includegraphics[scale=0.37]{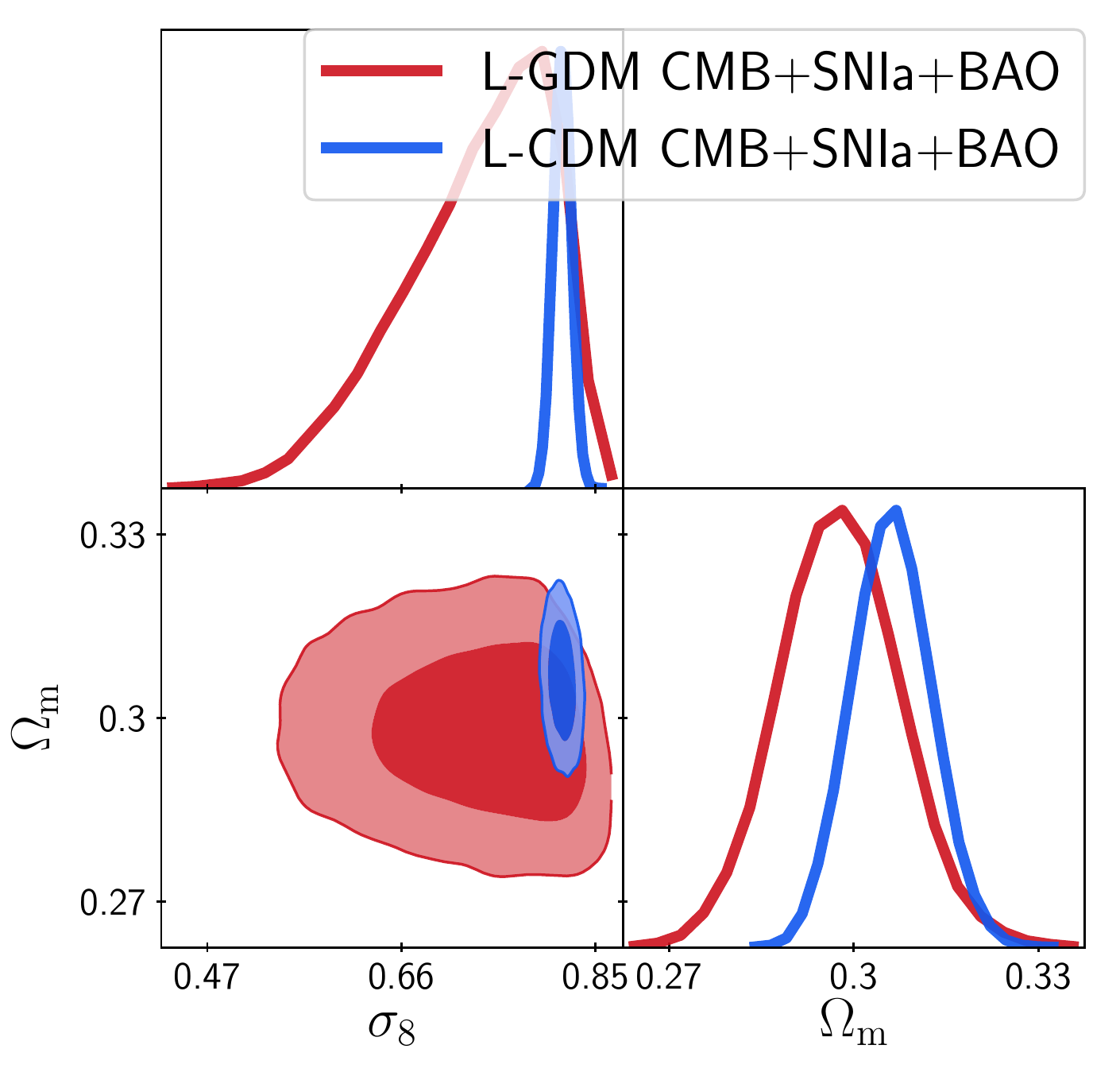}\hspace{30pt}\includegraphics[scale=0.37]{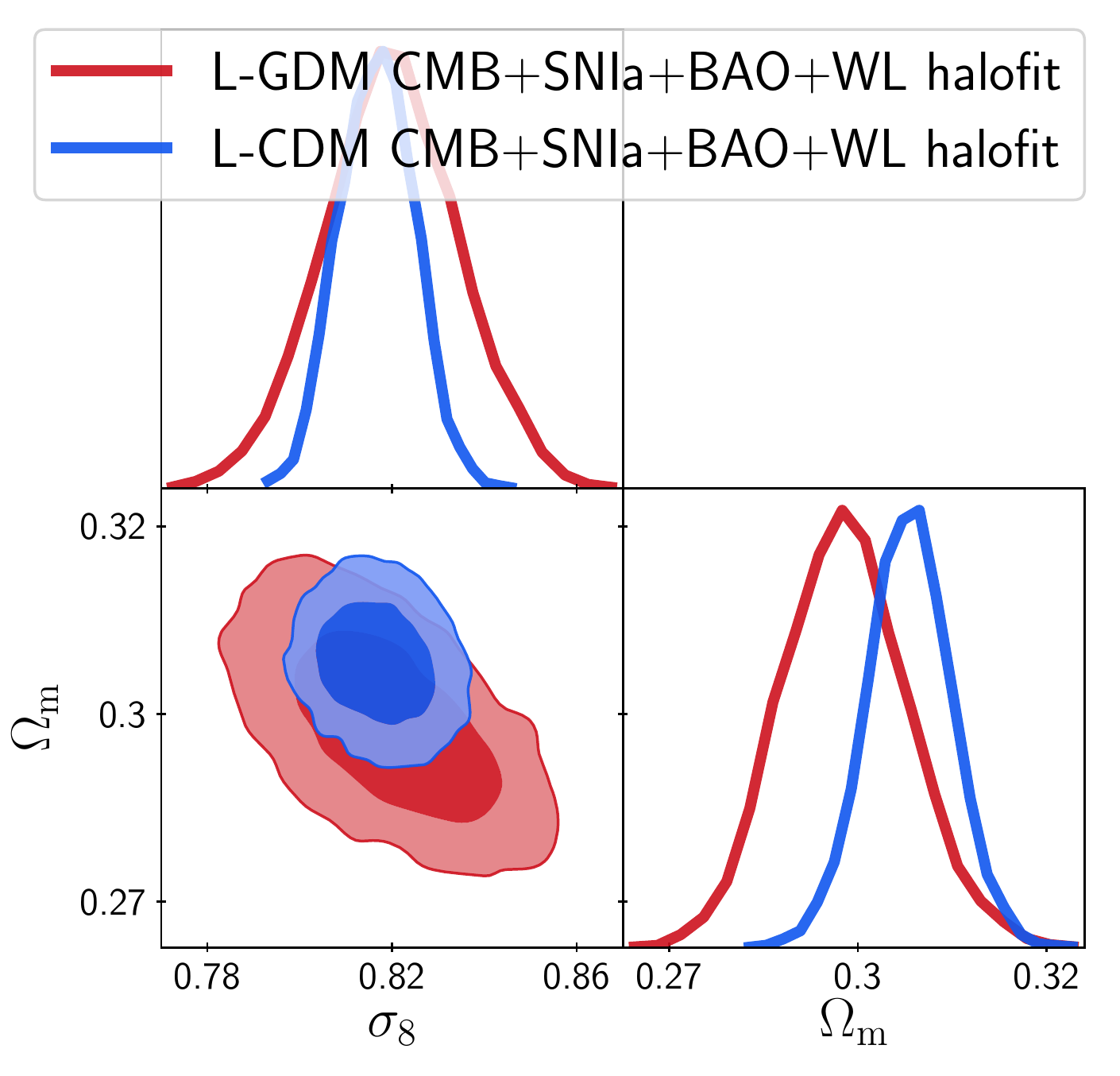}
\caption{1-$\sigma$ and 2-$\sigma$ contours for the $\Omega_{\rm m}$
  and $\sigma_8$ cosmological parameters for both $\Lambda$CDM (blue)
  and $\Lambda$GDM (red). {\it Left panel}: Combination of CMB, SNIa,
  and BAO data. {\it Right panel}: WL data are added into the analysis.}\label{fig1}
\end{center}
\end{figure}

%\subsection{Results}
In Table\,\ref{table1} we present the
constraints on the cosmological parameters when we fit both models,
$\Lambda$CDM and $\Lambda$GDM, to SNIa, BAO, and CMB data. The
constraints are weakened when we consider $\Lambda$GDM, due to the
introduction of two extra degrees of freedom. Even if all the parameters are
compatible within 1-$\sigma$ for both models, we can see that
$\Lambda$GDM allows for a smaller value of $\Omega_{\rm b}$,
$\Omega_{\rm dm}$ and $\sigma_8$, and a larger value of $h$
than $\Lambda$CDM. This points towards the fact that $\Lambda$GDM
could alleviate the tension between $\Omega_{\rm m}$ and the rms
matter density fluctuation $\sigma_8$
that appears when considering the $\Lambda$CDM model, as it can be
seen in the left panel of Fig.\,\ref{fig1}. Allowing for a non-vanishing sound speed strongly suppresses the matter
power spectrum at small scales, therefore, cosmological probes
sensitive to small scales are extremely important to constrain the GDM
parameters. In order to add WL data we need to take into account that
we enter the non-linear regime and our predictions for GDM are no
longer accurate. There is no non-linear recipe for GDM yet, so we use
an ultra-conservative approach by keeping only the largest scales from
CFHTLenS data, and we keep the standard halofit\,\cite{halofit}
non-linear correction. The constraints when WL data is added into the analysis
are shown in Table\,\ref{table1}. We can observe that the constraints
on most of the parameters (for both models) are equivalent to the ones obtained
without WL, since we are discarding most of the WL data. However, the
constraint on $c_s^2$ improves by
two orders of magnitude, due to the addition of information at mildly
non-linear scales. We need to remind, though, that the halofit
correction has been derived and tested only for standard cold dark matter, so the constraint on
$c_s^2$ could be slightly too optimistic. We can also see this effect in
the right panel of Fig.\,\ref{fig1}. The (probably)
over-estimated spectrum at small scales for $\Lambda$GDM gives very
good constraints on $\sigma_8$, questioning the ability of
$\Lambda$GDM to completely remove the tension between WL and CMB measurements.

\begin{table}[t]
\caption[]{Mean values with the 1-$\sigma$ constraints on the
  cosmological parameters for both $\Lambda$CDM and $\Lambda$GDM. Both
the constraints from the fit to CMB, SNIa, and BAO data (with and
without WL) and the
forecasted constraints from the photometric Euclid survey are shown.}
\label{table1}
\vspace{0.4cm}
\begin{center}
\small
%\resizebox{\textwidth}{!}{
\begin{tabular}{|c|c|c|c|c|}
\hline
Model & Parameters & CMB+SNIa+BAO & CMB+SNIa+BAO+WL & Photometric Euclid \\
\hline
 & $10^2\times \Omega_{\rm b}$ & $4.834\pm 0.054$ & $4.815\pm 0.053$ &
                                                                       $(4.834)\pm 0.075$\\
 & $\Omega_{\rm dm}$ & $0.2562\pm 0.0060$ & $0.2539\pm 0.0057$ & $(0.2562)\pm 0.0074$\\
 $\Lambda$CDM & $h$ & $0.6797\pm 0.0049$ & $0.6816\pm 0.0048$ & $(0.6797)\pm 0.016$\\
 & $\sigma_8$ & $0.8187\pm 0.0089$ & $0.8142\pm 0.0090$ & $(0.8187)\pm 0.0088$\\
 & $n_s$ & $0.9674\pm 0.0043$ & $0.9686\pm 0.0043$ & $(0.9674)\pm 0.023$\\
 & $10\times \tau$ & $0.72\pm 0.13$ & $0.68\pm 0.13$ & $-$\\
\hline

 & $10^2\times \Omega_{\rm b}$ & $4.71\pm 0.12$ & $4.69\pm 0.12$ &
                                                                       $(4.834)\pm 
                                                                       0.11$\\
 & $\Omega_{\rm dm}$ & $0.2491\pm 0.0087$ & $0.2458\pm 0.0088$ &
                                                                 $(0.2562)\pm 0.014$\\
 & $10^2\times w$ & $0.066\pm 0.054$ & $0.055\pm 0.053$ & $(0.00) \pm 0.21$\\
 $\Lambda$GDM & $10^6\times c_s^2$ & $<0.78$ &
                                                              $<0.010$ & $<0.0018$\\
& $h$ & $0.6873\pm 0.0082$ & $0.6898\pm 0.0082$ & $(0.6797)\pm 0.020$\\
 & $\sigma_8$ & $0.7351_{-0.041}^{+0.094}$ & $0.8174\pm 0.016$ &
                                                              $(0.8187)\pm 0.0096$\\
 & $n_s$ & $0.9656\pm 0.0044$ & $0.9682\pm 0.0042$ & $(0.9674)\pm 0.035$\\
 & $10\times \tau$ & $0.73\pm 0.15$ & $0.58\pm 0.16$ & $-$\\
\hline
\end{tabular}%}
\end{center}
\end{table}

\section{Euclid forecast}\label{sec_4}

In this section we focus our attention to study the $\Lambda$GDM model with
the future Euclid satellite\,\footnote{https://www.euclid-ec.org}
using the specifications of the Euclid Red-book\,\cite{Redbook}. We use the
\texttt{CosmoSIS}\,\cite{cosmosis} code\,\footnote{https://bitbucket.org/joezuntz/cosmosis/wiki/Home} to
compute a Fisher matrix forecast for the photometric Euclid
survey. In particular, for photometric galaxy clustering (GC), weak
lensing (WL), and their cross-correlations (XC). More in detail, we
replace (in the \texttt{CosmoSIS} pipeline) the standard Boltzmann code by our GDM modified version of \texttt{CLASS} used in
the previous section. %, and we also marginalize over several galaxy
                    %bias and intrinsic alignments nuisance
                    %parameters. 
We follow the previous approach of using the halofit correction and keeping only the largest
scales (up to $\ell_{\rm max}=750$). Concerning the fiducial cosmological model for the forecast, we use
the values obtained from the fit of $\Lambda$CDM to the combination of CMB,
SNIa, and BAO data from the previous section, including the treatment of massive neutrinos and
the value of $\tau$, which are fixed in the forecast. For the
parameters specific to GDM we consider the fiducial values $w=0$ and $c_s^2=10^{-9}$.

We can see in Table\,\ref{table1} that the photometric Euclid
survey will provide very good constraints on all parameters for both
models. For some parameters the forecasted constraints are worse than
the current ones. but this can be justified by the fact that Euclid
will only probe up to redshift $\sim 2.5$. Therefore, the lack of
high-redshift information coming from the CMB makes it harder to
constrain $w$ and break the degeneracies between the cosmological
parameters. We can infer that the combination of the full (photometric
and spectroscopic) Euclid data with the CMB will provide exquisite
constraints on GDM. It is worth mentioning that current low-redshift
probes (without the CMB) are only marginally able to constrain the GDM
parameters\,\cite{Tutusaus}. It is also important to notice that the
photometric Euclid survey alone will be able to constrain $c_s^2$
better than the combination of background and WL current data by
nearly an order of magnitude. However, as it was
the case in the previous section, we should treat these forecasted
constraints (especially on $c_s^2$) with
caution, since we know that the halofit correction is adapted to
standard cold dark matter.

\section{Conclusions}\label{conclusions}
In conclusion, we have seen that a more generalized treatment of dark
matter could alleviate the tension between low-redshift and
high-redshift data, thanks to a non-vanishing sound speed. Because of
$c_s^2$, the main differences between GDM and the
standard model appear at small scales. It is thus very important to
add cosmological probes sensitive to small scales to
constrain GDM. We have shown that adding WL
data strongly improves the constraints on the GDM sound speed. We have then focused on the photometric Euclid survey, and we have
shown that it will be able to put nice constraints on all parameters
(a very strong constraint on $c_s^2$), and it will allow us to increase our knowledge on the nature of
dark matter. However, it is necessary to have a non-linear recipe adapted
to GDM to be able to explore the small scales that Euclid will probe,
and extract the maximum of information of it.

%\section*{Acknowledgments}
%This work has been carried out thanks to the support of the OCEVU Labex (ANR-11-LABX- 0060) and of the Excellence Initiative of Aix-Marseille University - A*MIDEX, part of the French ``Investissements d'Avenir'' programme.

\end{document}